\documentstyle[11pt,psfig,draft]{nature}

\def\be{\begin{equation}}
\def\ee{\end{equation}}

\def\Zsun{{Z_\odot}}

\def\gsim{\lower.5ex\hbox{\gtsima}}
\def\lsim{\lower.5ex\hbox{\ltsima}}
\def\gtsima{$\; \buildrel > \over \sim \;$}
\def\ltsima{$\; \buildrel < \over \sim \;$}
\def\prosima{$\; \buildrel \propto \over \sim \;$}
\def\gsim{\lower.5ex\hbox{\gtsima}}
\def\lsim{\lower.5ex\hbox{\ltsima}}
\def\simgt{\lower.5ex\hbox{\gtsima}}
\def\simlt{\lower.5ex\hbox{\ltsima}}
\def\simpr{\lower.5ex\hbox{\prosima}}

\def\beq#1{\begin{equation}\label{#1}}
\def\eeq{\end{equation}}
\def\beqa#1{\begin{eqnarray}\label{#1}}
\def\eeqa{\end{eqnarray}}

\def\H2p{H$_2^+$ }

\def\mH2p{H_2^+}

\def\ApJ{{\it Astrophys. J. }}
\def\ApJL{{\it Astrophys. J. Lett. }}
\def\ApJS{{\it Astrophys. J. Suppl. Ser. }}
\def\MNRAS{{\it Mon. Not. R. Astron. Soc. }}
\def\AaA{{\it Astron. Astrophys. }}

\title{GRB 090423 at a redshift of $z\simeq 8.1$} 
\mainauthor{R. Salvaterra et al.}	

\author{{R.~Salvaterra\affiliation{INAF, Osservatorio Astronomico di Brera, via E. Bianchi 46, I-23807 Merate (LC), Italy}},
{M.~Della Valle\affiliation{INAF, Osservatorio Astronomico di Capodimonte, Salita Moiariello 16, 80131 Napoli, Italy}\affiliation{European Southern Observatory (ESO), 85748 Garching, Germany}\affiliation{International Centre for Relativistic Astrophysics, Piazzale della Repubblica 2, 65122 Pescara, Italy},
 {S.~Campana$^{~1}$},
 {G.~Chincarini\affiliation{Dipartimento di Fisica G.~Occhialini, Universit\`a di Milano Bicocca, Piazza della Scienza 3, I-20126 Milano, Italy}$^{~1}$},
 {S.~Covino$^{~1}$},
 {P.~D'Avanzo$^{~5 ~1}$},
 {A.~ Fern{\'a}ndez-Soto\affiliation{Instituto de Fisica de Cantabria, CSIC-Univ. Cantabria, Av. de los Castros s/n, E-39005 Santander, Spain}},
 {C.~Guidorzi\affiliation{Dipartimento di Fisica, Universita' di Ferrara, via Saragat 1, I-44100 Ferrara, Italy}},
 {F.~Mannucci\affiliation{INAF, Osservatorio Astrofisico di Arcetri, Largo E. Fermi 5, I-50125 Firenze, Italy}},
 {R.~Margutti$^{~5 ~1}$},
 {C.C. Th{\"o}ne$^{~1}$},
 {L.A.~Antonelli\affiliation{INAF, Osservatorio Astronomico di Roma, Via di Frascati 33, I-00040, Monte Porzio Catone, Rome, Italy}}},
 {S.D.~Barthelmy\affiliation{NASA,Goddard Space Flight Center, Greenbelt, MD 20771, USA}},
 {M.~De Pasquale \affiliation{Mullard Space Science Laboratory (UCL), Holmbury Rd, Holmbury St.Mary, Dorking, RH5 6NT, UK}},
 {V.~D'Elia$^{~9}$},
 {F.~Fiore$^{~9}$},
 {D.~Fugazza$^{~1}$},
 {L.K.~Hunt$^{~8}$}, 
 {E.~Maiorano\affiliation{INAF, IASF di Bologna, via Gobetti 101, I-40129 Bologna, Italy}},
 {S.~Marinoni\affiliation{INAF, Fundaci{\'o}n Galileo Galilei,
Rambla Jos{\'e} Ana Fern{\'a}ndez P{\'e}rez, 7 38712 Bre{\~n}a Baja, TF - Spain}\affiliation{Universit\'a degli Studi di Bologna, via Ranzani, 1, Bologna, Italy}},
 {F.E.~Marshall$^{~10}$},
 {E.~Molinari$^{~13~1}$}
 {J.~Nousek\affiliation{Department of Astronomy and  Astrophysics, Pennsylvania State University, University Park,  PA 16802, USA}},
 {E.~Pian\affiliation{INAF, Trieste Astronomical Observatory, Via G.B. Tiepolo 11, I-34143 Trieste, Italy}\affiliation{Scuola Normale Superiore, Piazza dei Cavalieri 1, 56100 Pisa, Italy}},
 {J.L.~Racusin$^{~15}$},
 {L.~Stella$^{~9}$},
 {L.~Amati$^{~12}$},
 {G.~Andreuzzi$^{~13}$},
 {G.~Cusumano\affiliation{INAF, Istituto di Astrofisica Spaziale e Fisica Cosmica di Palermo, Via Ugo La Malfa 153, 90146 Palermo, Italy}},
 {E.E. Fenimore}\affiliation{Los Alamos National Laboratory, P.O. Box 1663, Los Alamos, NM, 87545, USA},
 {P.~Ferrero\affiliation{Th{\"u}ringer Landessternwarte Tautenburg,  
Sternwarte 5, D-07778, Tautenburg, Germany}},
 {P.~Giommi\affiliation{ASI Science Data Center, ASDC c/o ESRIN, via G. Galilei, 00044 Frascati, Italy}},
 {D.~Guetta$^{~9}$},
 {S.T.~Holland$^{~10}$\affiliation{Universities Space Research Association, 10211 Wincopin Circle, Suite 500, Columbia, MD, 21044, USA}\affiliation{Centre for Research and Exploration in Space Science and Technology, Code 668.8, Greenbelt, MD, 20771, USA}},
 {K.~Hurley\affiliation{Space Sciences Laboratory, 7 Gauss Way, University of California, Berkeley, CA 94720-7450, USA}},
 {G.L.~Israel$^{~9}$},
 {J.~Mao$^{~1}$},
 {C.B.~Markwardt$^{~10~23}$\affiliation{Department of Astronomy, University of Maryland, College Park, MD 20742, USA}},
 {N.~Masetti$^{~12}$},
 {C.~Pagani$^{~15}$},
 {E.~Palazzi$^{~12}$},
 {D.M.~Palmer$^{~18}$},
 {S.~Piranomonte$^{~9}$},
 {G.~Tagliaferri$^{~1}$},
 {V.~Testa$^{~9}$}}

\begin{document}

\maketitle

\noindent
\rule{\textwidth}{0.5pt}

\noindent
{\bf Gamma-ray bursts (GRBSs) are 
produced by rare types of massive stellar explosions.  Their rapidly
fading afterglows are often bright enough at optical wavelengths, that
they are detectable  up to cosmological
distances. Hirtheto, the highest known redshift for a GRB
was $z=6.7$ (ref.~1), for GRB 080913, and for a galaxy was $z=6.96$
(ref.~2). Here we report observations of GRB~090423
and the near-infrared spectroscopic measurement of its redshift
$z=8.1^{+0.1}_{-0.3}$.
This burst happened 
when the Universe was only $\sim 4\%$ of its current age\cite{Komatsu09}.
Its properties are
similar to those of GRBs observed at low/intermediate redshifts,
suggesting that the mechanisms and progenitors that gave rise to this burst
about 600 million years after the Big Bang are not markedly
different from those producing GRBs $\sim 10$ billion years
later.}

\noindent
\rule{\textwidth}{0.5pt}

\bigskip

\noindent
GRB~090423 was detected by NASA's Swift satellite on 23 April 2009 at
07:55:19 UT as a double-peaked burst of duration $T_{90}=10.3\pm 1.1$ s. As
observed by Swift's Burst Alert Telescope
(BAT)\cite{Palmer09}, it had a 15--150 keV fluence 
$F=(5.9\pm0.4)\times 10^{-7}$ erg cm$^{-2}$ and a peak energy 
$E_p=48_{-5}^{+6}$ keV (errors at $90\%$ confidence level). 
Its X-ray afterglow was identified by Swift's X-ray Telescope (XRT), which
began observations 73 s after the BAT trigger\cite{Stratta09}. 
A prominent flare was detected at $t\sim 170$ s in the X-ray light
curve, which shows that a typical 'steep decay/plateau/normal decay'
behaviour (Fig.~1). Swift's UltraViolet Optical 
Telescope (UVOT) did not detect a counterpart even though  it started 
settled explosures
only 77 s after the trigger\cite{Pasquale09}. 
A 2$\mu$m counterpart
was detected with the United Kingdom Infra-Red Telescope (UKIRT, Hawaii) 20 min
after the trigger\cite{Tanvir09}.
Evidence that this burst occurred at high redshift, was
given by the multi-band imager Gamma-Ray Burst Optical/Near-Infrared
Detector (GROND, Chile) multiband imager (from $g$' band to $K$ band), which 
indicated a photometric
redshift of $z = 8.0^{+0.4}_{-0.8}$ (ref.~7).

We used the 3.6m Telescopio Nazionale Galileo (TNG, La Palma) with the  Near Infrared 
Camera Spectrometer (NICS)  
and the Amici prism to obtain a low-resolution ($R\approx50$) spectrum of 
GRB~090423 $\sim 14$ hrs after the trigger. NICS/Amici 
is an ideal instrument to detect spectral breaks in the 
continuum of faint objects  because of its high efficiency and 
wide simultaneous spectral coverage (0.8-2.4 $\mu$m). 
The spectrum (Fig.~2) reveals a clear break at a wavelength of 1.1 $\mu$m 
(ref.~8). 
We derive
a spectroscopic redshift for the GRB of $z = 8.1^{+0.1}_{-0.3}$ (ref.~9; see 
Supplementary Information, section~3), interpreting  
the break as Lyman-$\alpha$ absorption in the intergalactic medium.  No other
significant absorption features were detected.
This result is consistent, within the errors, with the measurement reported in
ref.~7.  

At $z\sim 8.1$, GRB~090423 has a prompt-emission rest-frame duration
of only $T_{90,rf}=1.13\pm0.12$ s in
the redshifted 15-150 keV energy band, an isotropic equivalent energy
$E_{iso}=1.0\pm0.3\times 10^{53}$ erg in the redshifted 8-1000 keV energy 
band\cite{Kienlin09} and a peak energy
$E_{p,rf}=437\pm55$ keV.  The short duration and the high peak energy
are consistent both with the
distribution of long bursts, linked to massive stellar collapse,
and with the population of short bursts, thought  to arise from
the merger of binary compact stars\cite{Meszaros,Zhang}. 
Although the analysis of the spectral lag
between the high- and low-energy channels in the BAT band is inconclusive about
the classification of GRB~090423, the high $E_{iso}$ argues in favor of a
long GRB. The fact that GRB~090423 matches the
$E_{iso}-E_{p,rf}$ correlation of long GRBs within 0.5$\sigma$ further
supports this classification\cite{Amati08} (Supplementary Fig.~2). 

The rest-frame $\gamma-$ray and X-ray light curve of GRB~090423 is remarkably
akin to those of long GRBs at low, intermediate and high redshifts (Fig.~1),
suggesting similar physics and  
interaction with the circumburst medium. The near-infrared light curve
of GRB~090423 $\sim 15$h after the trigger shows a temporal decay with
a power-law index of $\alpha_0\sim 0.5$, which is
markedly different from the decay observed at X-ray energies during the 
same time interval, which has a power-law index of $\alpha_{X,2}\sim
1.3$ (Supplementary Fig.~3 and Supplementary Information, section~2). 
As for other lower-redshift GRBs, this 
behaviour is difficult to reconcile with standard afterglow models,   
although the sampling of the near-infrared light curve is too
sparse for any firm conclusion to be drawn.

The spectral energy distribution of near-infrared afterglow is well fitted
by a power-law with an index of $\beta = 0.4^{+0.2}_{-1.4}$
and an equivalent interstella extinction of E($B-V$)$<0.15$, assuming 
dust reddening consistent with the Small Magellanic Cloud\cite{Soto09}.
On the other hand, the analysis of the XRT data in the
time interval 3900s--21568s suggests the presence of intrinsic
absorption (in excess of the Galactic value) with
an equivalent hydrogen column density of
$N_H(z)=6.8^{+5.6}_{-5.3}\times 10^{22}$ cm$^{-2}$ (90\% confidence
level; Supplementary Information, section~1). The low value of the dust
extinction coupled with a relatively high value of $N_H$ suggests that GRB~090423
originates from a region with low dust content relative to those of low-$z$
GRBs\cite{Schady07}, but one similar to that of the high-$z$ GRB 050904,
for which $z=6.3$ (ref.~15).
Because the absorbing medium must be thin
from the point of view of ``Thomson'' scattering, 
the metallicity of the circumburst medium can be
constrained to be $>4$\% of the solar value, $\Zsun$. 
The implication is  that previous
supernova explosions have already enriched the host galaxy of
GRB~090423 to more than the critical metallicity, $Z\sim
10^{-4}\;\Zsun$ (ref.~16) that prevents the formation of very
massive stars (population III stars). Therefore the
progenitor of GRB~090423 should belong to a second stellar generation. 
Its explosion injected fresh metals into the interstellar medium, further
contributing to the enrichment of its host galaxy. Its existence
empirically supports the  
cosmological models\cite{Springel05,Nagamine06} in which
stars and galaxies, already enriched by metals, are in place only
$\sim 600$ million years after the Big Bang. Long GRBs
are mostly associated with star forming dwarf galaxies, which are thought to be 
the dominant population of galaxies in the early
Universe\cite{Choudhury08}. The fact that GRB~090423 appears
to have exploded in an environment similar to that of low-$z$ GRB 
hosts\cite{Fruchter} is in agreement with this. 

The occurrence of a GRB at $z\sim 8$ has important implications for the cosmic
history of these objects\cite{Lamb,Guetta05,Bromm06,SC07}.
In a first, simple approach, we can assume that GRBs trace the cosmic star formation 
history, given the well-known link of the long GRBs and the
deaths of massive stars\cite{WB}, and that GRBs are
well described by a universal luminosity function. However, under 
these assumptions the expected number of
bursts at $z\ge 8$ with an observed photon peak flux larger than or equal to that of 
GRB~090423 is extremely low: $\sim 4\times 10^{-4}$ in 
$\sim 4$ yrs of Swift operation (Supplementary Fig.~6 and Supplementary
Information, section~4). 
Hence, one or both of the above assumptions may be oversimplification\cite{SC07,RS08}.
The detection of a very high-$z$ burst such as GRB 090423 could be accomodated
if the GRB luminosity function were shifted towards
higher luminosity according to $(1+z)^\delta$ with $\delta\simgt 1.5$
or if the GRB
formation rate were strongly enhanced in galaxies with $Z\simlt
0.2\Zsun$. 
The requirement for evolution may be mitigate if we assume a very high
star formation rate at $z>8$. However, we note that the need for
evolution is strongly supported by both the large number of Swift
detections at $z>2.5$ (ref.~24) and the number of bursts with
peak luminosities in excess of $10^{53}$ erg s$^{-1}$ (ref.~26).
A possible explanation is that high-redshift galaxies are 
characterized by a top-heavy (bottom-light)
stellar initial mass function with a higher incidence of massive
stars than in the local Universe\cite{chary}, providing an 
enhanced number of GRB progenitors. Such objects could be the main agents 
responsible for completing the reionization of the 
Universe\cite{Bolton07,Choudhury08,Furlanetto09,Stiavelli}.

\noindent
\large{\bf Acknowledgements} We acknowledge the TNG staff for useful
support during ToO observations, in particular A. Fiorenzano,
N. Sacchi, A.G. de Gurtubai Escudero. 
We thank A.~Ferrara for useful discussions. 
This research was supported by the
Agenzia Spaziale Italiana, the Ministero dell'Universit{\`a} e della
Ricerca (MUR), the Ministero degli Affari Esteri, NASA, and 
the National Science Foundation (NSF).

\noindent
All authors made contributions to this paper. This took the form of
direct analysis of the Swift data (SC, GC, CG, RM, SDB, MDP, FEM, JN,
JLR, GC, EEF, PG, STH, JM, CBM, CP, DMP), analysis of the TNG and
photometric data (MdV, SC, PDA, AFS, CCT, LAA, FM, VE, FF, DF, LKH,
EM, EM, SM), management of optical follow-up (PDA, LAA, VDE, EM, SM,
GA, PF, GLI, NM, EP, SP, GT, VT),
interpretation of the GRB properties (RS, MdV, SC, GC, SC, PDA, AFS,
CG, RM, CCT, LA, EP, LS, KH), and modeling of the GRB luminosity function (RS,
MdV, SC, GC, CG, DG GT). Additionally, all authors have made
contributions through their major involvement in the programmes from
which the data derives, and in contributions to the interpretation,
content and discussion presented here.

\noindent
\large{\bf Competing interests statement} The authors declare that they
have no competing financial interests. 

\noindent
\large{\bf Correspondence} and requests for materials should be addressed to R.S.
(e-mail: salvaterra@mib.infn.it).

\newpage

\begin{figure*}[t]
\begin{center}
\centerline{\psfig{figure=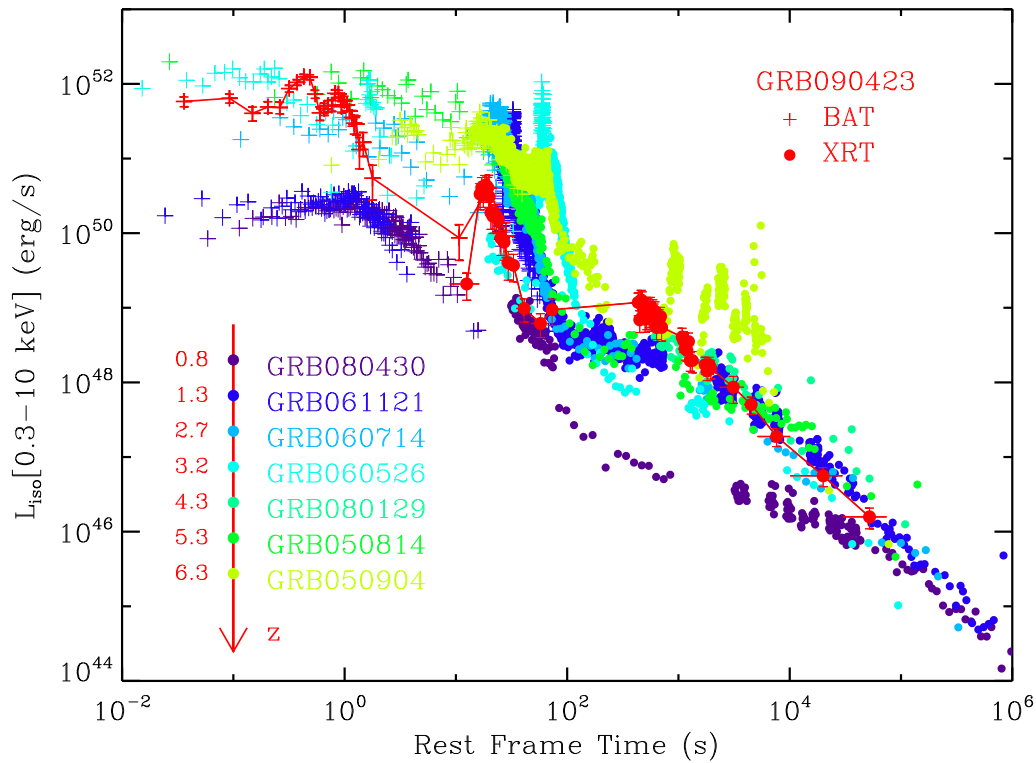,height=11cm}}
\caption{{\bf Rest-frame $\gamma$-ray and X-ray light curves for bursts at different 
redshifts.} {\footnotesize 
BAT and XRT light curve of GRB~090423 (red data) in the source
rest-frame. Errors on luminosity, $L_{iso}$, are at 1$\sigma$ level; horizontal bars 
refer to the integration time interval.
The XRT 0.3--10 keV light-curve shows a prominent flare at a rest-frame time of 
$t_{rf} \sim 18$ s (also detected by BAT), and a flat phase
(with a power-law index of $\alpha_{X,1}=0.13 \pm 0.11$) followed by a rather typical 
decay with power-law index $\alpha_{X,2}=1.3\pm0.1$. We compare the light curve
of GRB~090423 with those of seven GRBs in the redshift interval 0.8-6.3. The bursts are
selected from among those showing a canonical three-phase behaviour 
(steep decay/plateau/normal decay) in the
X-ray light curve  and without a spectral break between BAT and XRT,
allowing the spectral calibration of the BAT signal into the 0.3-10
keV energy band. The light curve of GRB~090423
does not show any distinguishing features relative to those of the lower-redshift bursts,
suggesting that the physical mechanism that causes the GRB and its
interaction with the circumburst medium are similar at $z\sim 8.1$ and at
lower redshifts.}
}
\label{fig3}
\end{center}
\end{figure*}

\begin{figure*}[t]
\begin{center}
\centerline{\psfig{figure=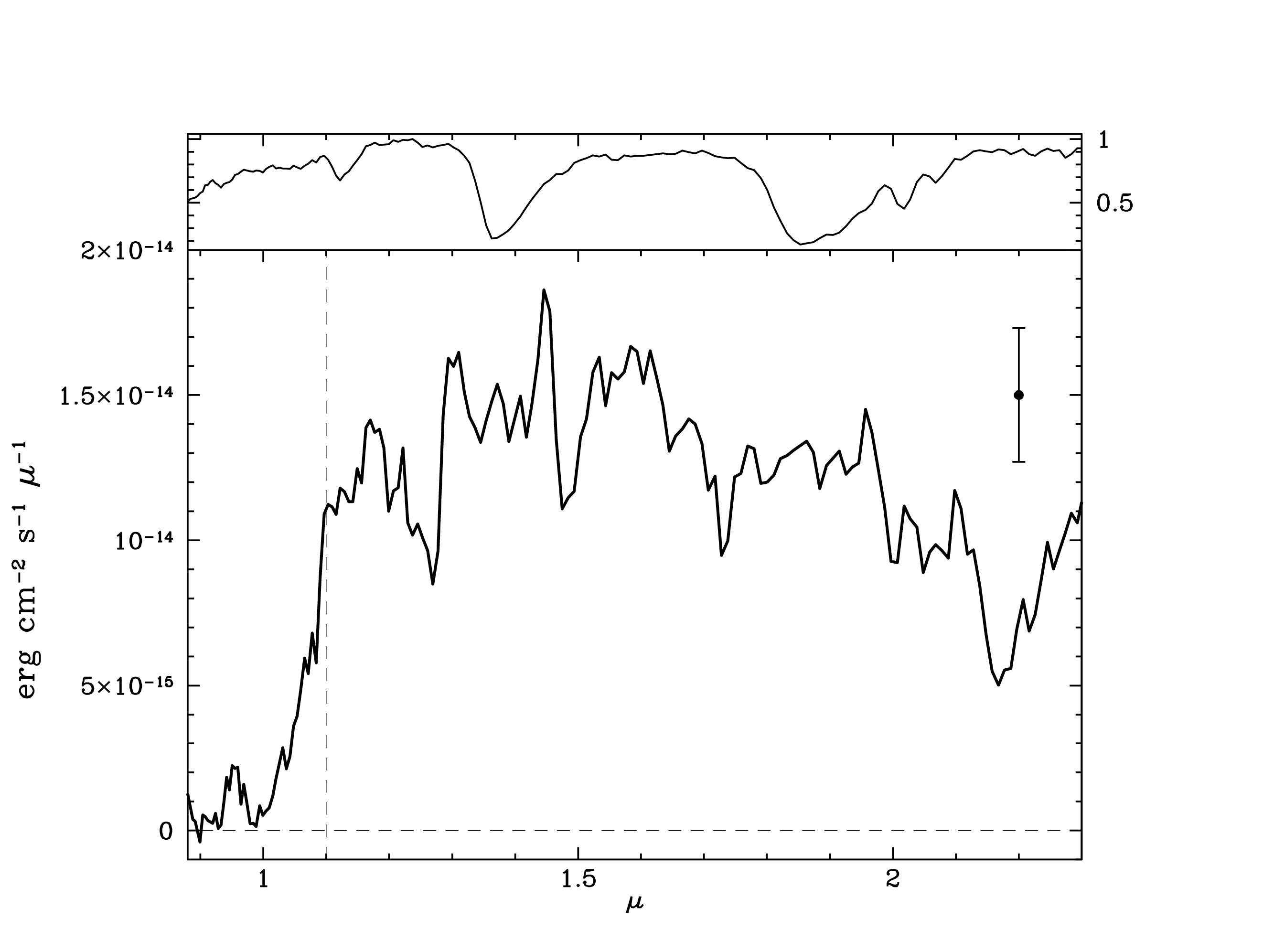,angle=-90,height=11cm}}
\caption{{\bf TNG spectrum of the NIR afterglow.} {\footnotesize Bottom Panel. 
Spectrum of GRB 090423 obtained using the Amici prism on the Telescopio Nazionale 
Galileo (TNG). The sharp break at wavelength $\lambda \approx 1.1$ $\mu$m, 
which is due to the HI 
absorption in the intergalactic medium at the wavelength of the Ly$\alpha$ line, 
implies that $z=8.1^{+0.1}_{-0.3}$. The spectrum has been smoothed with a boxcar filter 
of width $\Delta=25$ pixels (where one pixel corresponds to $\sim 0.006$ $\mu$m at 
$\lambda=1.1\mu$m).  The absolute flux calibration was obtained by matching the 
almost simultaneous GROND photometric measurements\cite{Tanvir09}. The wavelength 
calibration was obtained from the TNG archive and adjusted to the wavelengths of the 
main atmospheric bands.  The error bar corresponds to $\pm 1\sigma$ uncertainty as 
measured on the smoothed spectrum. The confidence level of the Lyman-$\alpha$ break 
detection is $\simgt 4\sigma$. See also Supplementary Information, section~3.
Top Panel. Plot of the trasmittance (the atmospheric transparency convolved with the 
instrumental response). The system has a significant sensitivity down to 0.9 $\mu$m, 
and no instrumental or atmospheric effect could explain the abrupt flux break observed 
in the spectrum of GRB~090423.}}
\label{fig2}
\end{center}
\end{figure*}


\newpage
\clearpage
\normalsize

{\bf Supplementary Information}

{\bf This material presents technical details to support the discussion in 
the main paper. We discuss here the Swift data analysis, the analysis of the photometric data, the details of the analysis of the TNG spectroscopic data
and the modelling of the GRB redshift distribution.}

\bigskip

\noindent
{\bf 1. Swift data analysis}

Swift-BAT triggered on GRB~090423 at 07:55:19.35 UT on 23 April
2009. 
BAT data were analysed using the HEASOFT software package (version 6.6.2)
with the Swift Calibration Database (CALDB) version
BAT(20090130). Background-subtracted light curves in different energy
channels, energy spectra and corresponding response functions were
derived from the BAT event file as processed with the BAT software tool {\tt
batgrbproduct}, by using the mask-weighting technique for the BAT refined 
position$^4$, and by
using standard and BAT-dedicated software tools.  The
mask-weighted light curve (Supplementary Figure 1) showed a couple of
overlapping peaks starting at $T_0-2$ s, peaking at $T_0+4$ s, and
ending at $T_0+15$ s. The estimated duration, $T_{90}$, was $10.3\pm1.1$ s 
for the mask-weighted light curve in the 15-150 keV band. $T_{90}$ is
defined as the duration 90\% of the total prompt $\gamma$-ray fluence in
the observer frame (i.e. the interval from 5\% to 95\% of the total fluence)
and is estimated using the {\tt battblocks} software tool. 
Noteworthy is the light curve of the hardest channel, from 100
to 150 keV, showing a very weak signal as compared with those of the
other energy channels. This reflects the spectral softness of this GRB, as shown also
by the total energy spectrum. The latter was accumulated from -0.7 to
11.7 s and is fitted with a cut-off power law, $N(E)\sim E^{-\Gamma}
\exp[(2-\Gamma)E/E_p]$, with the best-fit value for the peak energy, 
$E_p=48_{-5}^{+6}$ keV, and a photon index $\Gamma=0.6_{-0.6}^{+0.5}$. 
This value of $E_p$ is fully consistent with that determined by fitting the 
Fermi/Gamma-ray Burst Monitor spectrum$^{10}$ with the canonical
"Band" function\cite{Band93}.
The corresponding total fluence in the 15-150 keV energy band is $(5.9\pm
0.4)\times 10^{-7}$ erg cm$^{-2}$. The 1-s peak photon flux measured
from 3.5 s in the 15-150 keV band is $1.7\pm0.2$ ph s$^{-1}$ cm$^{-2}$.
 Uncertainties are given at 
90\% confidence. At $z=8.1$, GRB~090423 is found to be consistent with
the  $E_{\rm p,rf}$ -- $E_{iso}$ correlation$^{13}$ within 0.5$\sigma$ 
(Supplementary Figure 2). We note that, even considering the measured peak energy as obtained
by a fit of the Fermi/GBM data with a cut-off power-law spectrum,
i.e. $E_p = 82\pm 15$ keV and thus $E_{p,rf} = 746\pm 137$ keV
(ref. $^{10}$), GRB 090423 would still be consistent within
2$\sigma$ with the $E_{p,rf}-E_{iso}$ correlation.

The BAT light curve shown in Supplementary Figure 3 is the mask-weighted curve
extracted between 15 and 150 keV, binned so as to ensure S/N$>2$ with a
minimum binning time of 0.512 s. 
Extrapolation of the BAT flux
down to the 0.3-10 keV band was performed by assuming the above spectral model.
We note that the remarkable X-ray flare detected by XRT is
seen in the BAT data as well. We note that assuming that the flare is still
part of the prompt emission\cite{Zhang09}, the total
duration of the prompt phase in the source rest frame might be $\sim 20$s, 
similar to other long GRBs.

\bigskip
The XRT observations began 73 s after the trigger: up to $\sim 300$ s  
the signal was dominated by a flare. As in many other GRBs,
the light-curve then flattened to a shallow decay phase which could be
well modelled by a power-law with  index $\alpha_{X,1}=0.13\pm0.11$.
At $t\sim 4500$ s the X-ray afterglow steepened to $\alpha_{X,2}=1.3\pm0.1$
(errors at 68\% confidence level).
The flare was modelled by a standard profile\cite{Norris05}: this is
characterised by a $1/e$ rise-time $t_{rise}=29.1\pm3.6$ s; $1/e$ decay-time
$t_{decay}= 65.5\pm3.6$ s; $1/e$ width of $\Delta t=94.6\pm 7.3$ s, while the 
asymmetry parameter is $k=0.38\pm0.03$. This implies a variability 
measure $\Delta t/t_{peak}=0.66$ and a brightness contrast $\Delta$Flux/Flux
around 25. While the flare parameters are defined following ref.~\cite{Norris05},
the reported uncertainties are worked out by using the entire
covariance matrix. At the redshift of the burst, the flare has an energy 
$E_{iso}=3.6\times 10^{51}$ erg in the redshifted 0.3-10 keV band of XRT, 
comparable to the energy released during the prompt emission of other GRBs.

To evaluate the intrinsic column density absorbing the GRB~090423
spectrum, we extracted data in the 3900--21568 s time interval
(observer frame). 
This interval was selected in order to avoid the bright X-ray flare
whose variable spectrum might alter the fit and in order to have
sufficient signal in the extraction region which we define as a
count rate of more than 0.01 counts s$^{-1 }$.
The resulting 7984 s exposure contains 680 counts
in the range between 0.3-10 keV.  The ancillary response file (arf) was 
created with the task {\tt xrtmkarf} (within {\tt heasoft} v.6.2.2) using the 
relevant exposure file and the latest v.11 reponse matrix function (rmf). 
The spectrum was binned to 20 counts per bin in order to assure a 
reasonable $\chi^2$ statistic.

We fit the X-ray spectrum with a composite absorption model consisting of a 
Galactic contribution and an intrinsic absorption fixed to z=8.1 using the 
{\tt tbabs} model within the XSPEC (v12.5.0aa) package.
We left the Galactic value free to vary in the
$2.9-3.2\times 10^{20}$ cm$^{-2}$ range (based on the absorption maps
by \cite{Dickey90} and \cite{Kalberla04}).
The X-ray continuum was modeled with a power law, as is customary for
the afterglow spectra of GRBs. The overall fit is good
with a reduced $\chi^2_{\rm red}=1.12$ (28 degrees of freedom,
corresponding to a null hypothesis probability of $30\%$). The
resulting power law photon index is $\Gamma_X=1.97^{+0.15}_{-0.16}$. For
the intrinsic column density, we get a value of
$N_H(z)=6.8^{+5.6}_{-5.3}\times10^{22}$ cm$^{-2}$ (90\% confidence level), 
among the highest of all Swift GRBs\cite{Evans09}.  
The results refer to a solar composition and metallicity.
Assuming that the medium is not Thomson thick, a lower limit of the
metallicity can be obtained by $N_H(z)\simlt (1/\sigma_T) (Z/Z_\odot)^{-1}$, where
$\sigma_T$ is the Thomson cross-section (e.g. ref. \cite{Campana07}).
We find $Z>0.043\;Z_\odot$.
A lower limit on the
value of $N_H(z)>6\times 10^{21}$ cm$^{-2}$ is found at 95\% confidence level
corresponding to a lower limit on the metallicity of $Z>0.004\;Z_\odot$.

\bigskip

\noindent
{\bf 2. Analysis of the photometric data}

We analyzed all the available photometric data$^{7}$ by using the {/it zphrem} code\ref{Soto03},
in order to determine the photometric redshift and spectral properties
of the afterglow. Our code fits a model of functional form $f_\nu \propto
\nu^{-\beta} t^{-\alpha_O}$, including dust extinction (by an SMC-type
extinction law) as a free parameter.

We find that the data are best fit within the time range $4.2 \times 103 <
t < 6.6 \times 104$ s) with a model characterised by a temporal decay with
a power law index $\alpha_O = 0.50 \pm 0.05$ (we quote hereafter 95\%
confidence intervals). The dust content is constrained to be $E(B-V)<
0.15$, and the photometric redshift is $z_{\rm phot} = 8.3 \pm 0.3$,
consistent with the spectroscopic results. We caution that the rest-frame
wavelength observed extends only out to 2500\AA, and only the three $JHK$
filters do indeed measure any flux redwards of Lyman-$\alpha$. That is the
reason why the spectral index is only loosely constrained and its error is
asymmetric ($\beta = 0.4^{+0.2}_{-1.4}$), although its relatively blue color
still enables us to put stringent limits to the possible dust content in
the afterglow environment. Supplementary Figure 4 shows the projections of
the $(z, \alpha_O, \beta, E(B-V))$ four-dimensional confidence intervals on
the different bidimensional planes.

Extending the analysis to the whole available temporal window ($2.5 \times
10^2 < t < 1.4 \times 10^6$ s) renders impossible to find a good fit with a
single temporal power-law, because of the different decay regimes that the
afterglow goes through.

\bigskip

\noindent
{\bf 3. Analysis of the TNG spectroscopic data}

We observed the afterglow of GRB090423 with the near-IR camera and
spectrograph NICS\cite{Baffa01} on the Italian 3.6m Telescopio
Nazionale Galileo (TNG) at La Palma.  We used the the lowest spectral
resolution 
mode, offered by the Amici prism\cite{Oliva00}. This prism provides a
simultaneous spectral coverage over a wide wavelength range, between
0.8 and 2.4 $\mu$m, and has a high efficiency.  It yields a constant
spectral resolution $R \approx 50$ over the whole wavelength range. 
These characteristics make the instrument especially well-suited for
studying the spectral distribution of faint objects.

We obtained 128 minutes of on-target spectroscopy. The afterglow was
positioned in the slit using as reference a nearby star approximately
30 arcseconds away (at J2000 coordinates 09:55:35.31, +18:09:03.9). We
used a dithering mosaic of 8 cycles, each including two coadds of
single 120s exposures, repeated 4 times. The mean time of our
observations was Apr 23.98, approximately 15.5 hours after the burst
detection. The 2-dimensional spectrum is shown in Supplementary Figure 5.

Standard reduction tasks for NIR spectroscopy were performed
independently by four different groups in our team, all of them
reaching consistent results. Wavelength calibration was obtained by
using a standard calibration table provided by the TNG and 
matching the deep telluric absorption bands. This method allows for 
wavelength calibrations better than 0.005 $\mu$m at 1.1 $\mu$m,
and its contribution ($\Delta z= \pm$0.04) to the final error budget on 
redshift is negligible.

Relative flux calibration was performed by using the observed
spectral shape of the reference star. Its optical (SDSS) and near-IR
(2MASS) colors are consistent with those of an M3-III star. The
absolute calibration of the spectrum was obtained from the
comparison with the simultaneous photometric measurements obtained by
GROND (H=19.94 (Vega), ref. $^7$). We estimate the
slit losses to be less than 30\%.

The observed flux is compatible with zero below a wavelength of 1.1 $\mu$m,
while a significant flux ($>99\%$ confidence level) is measured redwards
of this limit. Assuming that this is due to hydrogen absorption by a
virtually completely thick Lyman-$\alpha$ forest, then the redshift at
which the GRB occurred is $z= 8.1^{+0.1}_{-0.3}$. The quoted error
includes the uncertainties on the wavelength calibration and on the
estimate of the break position. This
value makes GRB090423 the most distant object spectroscopically
identified to date. By using a standard cosmology with
$\Omega_\Lambda = 0.73$, $\Omega_M = 0.27, H_0=71\, {\rm km s}^{-1}
{\rm Mpc}^{-1}$, we find that GRB~090423 was detected at a lookback
time of greater than 13 Gyrs.

We tentatively identified two absorption features at 1.3 and 2.2
$\mu$m. These would be consistent with blends of Si IV and Fe II at
1400\AA~ and 2400\AA, $z=8.1$ rest-frame, respectively. The detection,
however, has a low confidence level due to the low S/N of the
spectrum.

\bigskip

\noindent
{\bf 4. Modelling the GRB redshift distribution}

We compute the probability of detecting of GRB~090423 in three
different scenarios for the formation and cosmic evolution of long
GRBs: (i) no evolution model, where GRBs follow the cosmic star
formation and their luminosity function (LF) is constant in redshift; 
(ii) luminosity
evolution model, where GRBs follow the cosmic star formation but the
LF varies with redshift; (iii) density evolution model, where GRBs
form preferentially in low--metallicity environments. In the first two
cases, the GRB formation rate is simply proportional to the global
cosmic star formation rate as computed by \cite{Hopkins}. 
For the luminosity evolution model, the
typical burst luminosity is assumed to increase with redshift as
$(1+z)^\delta$.  Finally, for the density evolution case, the GRB
formation rate is obtained by convolving the observed SFR with the
fraction of galaxies at redshift $z$ with metallicity below $Z_{th}$
using the expression computed by \cite{Langer06}. In this
scenario, the LF is assumed to be constant.

The computation works as follows (see also $^{21,22,23,24,26,43}$).
The observed photon flux, $P$, in the energy band 
$E_{\rm min}<E<E_{\rm max}$, emitted by an isotropically radiating source 
at redshift $z$ is

\begin{equation}
P=\frac{(1+z)\int^{(1+z)E_{\rm max}}_{(1+z)E_{\rm min}} S(E) dE}{4\pi d_L^2(z)},
\end{equation}

\noindent
where $S(E)$ is the differential rest--frame photon luminosity of the source, 
and $d_L(z)$ is the luminosity distance. 
To describe the typical burst spectrum we adopt the
functional form proposed by \cite{Band93}, i.e. a broken power--law
with a low--energy spectral index $\alpha$, a high--energy spectral index
$\beta$, and a break energy $E_b=(\alpha-\beta)E_p/(2+\alpha)$, 
with $\alpha=-1$ and $\beta=-2.25$ (ref. \cite{Preece00}).
In order to broadly estimate the peak energy of the spectrum, $E_p$, 
for a given isotropic--equivalent peak luminosity, 
$L=\int^{10000\,\rm{keV}}_{1\,\rm{keV}} E S(E)dE$, we assumed
the validity of the correlation between $E_p$ and $L$ (ref. \cite{Yonetoku04}). 

Given a normalized GRB LF, $\phi(L)$, the observed rate of 
bursts with $P_1<P<P_2$ is

\begin{equation}
\frac{dN}{dt}(P_1<P<P_2)=\int_0^{\infty} dz \frac{dV(z)}{dz}
\frac{\Delta \Omega_s}{4\pi} \frac{\Psi_{\rm GRB}(z)}{1+z} \int^{L(P_2,z)}_{L(P_1,z)} dL^\prime \phi(L^\prime),
\end{equation}

\noindent
where $dV(z)/dz$ is the comoving volume element,
$\Delta \Omega_s$ is the solid angle covered on the sky by the survey,
and the factor $(1+z)^{-1}$ accounts for cosmological time dilation. 
$\Psi_{\rm GRB}(z)$ is the comoving burst formation rate and
the GRB LF is described by a power law with an exponential
cut--off at low luminosities\cite{PM01}, i.e. $\phi(L) \propto 
(L/L_{\rm cut})^{-\xi} \exp (-L_{\rm cut}/L)$.

For the three scenarios, we optimize the model free parameters (GRB
formation efficiency, burst typical luminosity at $z=0$ and the power index
$\xi$ of the LF) by fitting the differential number counts observed by BATSE
(see ref. $^{24,26}$ for a
detailed description of the models and of the analysis). We find that
it is always possible to find a good agreement between models and
data. Moreover, we can reproduce also the differential peak
flux count distribution observed by Swift in the 15-150 keV band without
changing the best fit parameters. On the basis of these results, we
compute the probability to detect with Swift a GRB at $z\ge 8$ with
photon flux $P$. The results are plotted in Supplementary Figure 6 (top panels)
together with the cumulative number of GRBs at $z\ge 8$ expected to
be detected by Swift in one year of observations (bottom panels). From
the plot it is clear that the no evolution model fails to account for
the observation of GRB~090423, since only $\sim 4\times 10^{-4}$ GRBs are expected 
to be detected  at $z\ge 8$ in $\sim 4$ years of Swift observations.
Evolutionary models (both in luminosity or in density) can easily
account for the discovery of GRB~090423. 
We note that the results confirm the need for cosmic
evolution in the GRB luminosity function and/or in the GRB density
obtained by recent analysis of the whole Swift GRB dataset. Indeed, both the
large number of $z\ge 2.5$ bursts$^{24}$ and the number
of bright (i.e. with peak luminosity $L\ge 10^{53}$ erg s$^{-1}$)  
bursts$^{26}$ strongly require the existence of evolution.

Moreover, we want to stress here that our conclusions are conservative. First 
of all, many biases can hampered the detection of GRB at very high redshift. 
Indeed, a few very high-$z$ bursts may be hidden among the large sample of 
Swift bursts that lack of an optical detection. Thus, the discovery of a single
event at $z>8$ in 4.5 yrs of Swift operation can be treated as a lower limit
on the real number of high-$z$ detection. Moreover, our choice of the GRB LF 
is also  conservative, since the existence of large population of faint GRBs
(i.e. for an LF with a more gentle decline or a rise in the faint end) would 
lead to a decrease of the expeceted number of GRBs at $z>8$ strenghtening our 
conclusions.

\newpage

\begin{figure*}[t]
\begin{center}
\centerline{\psfig{figure=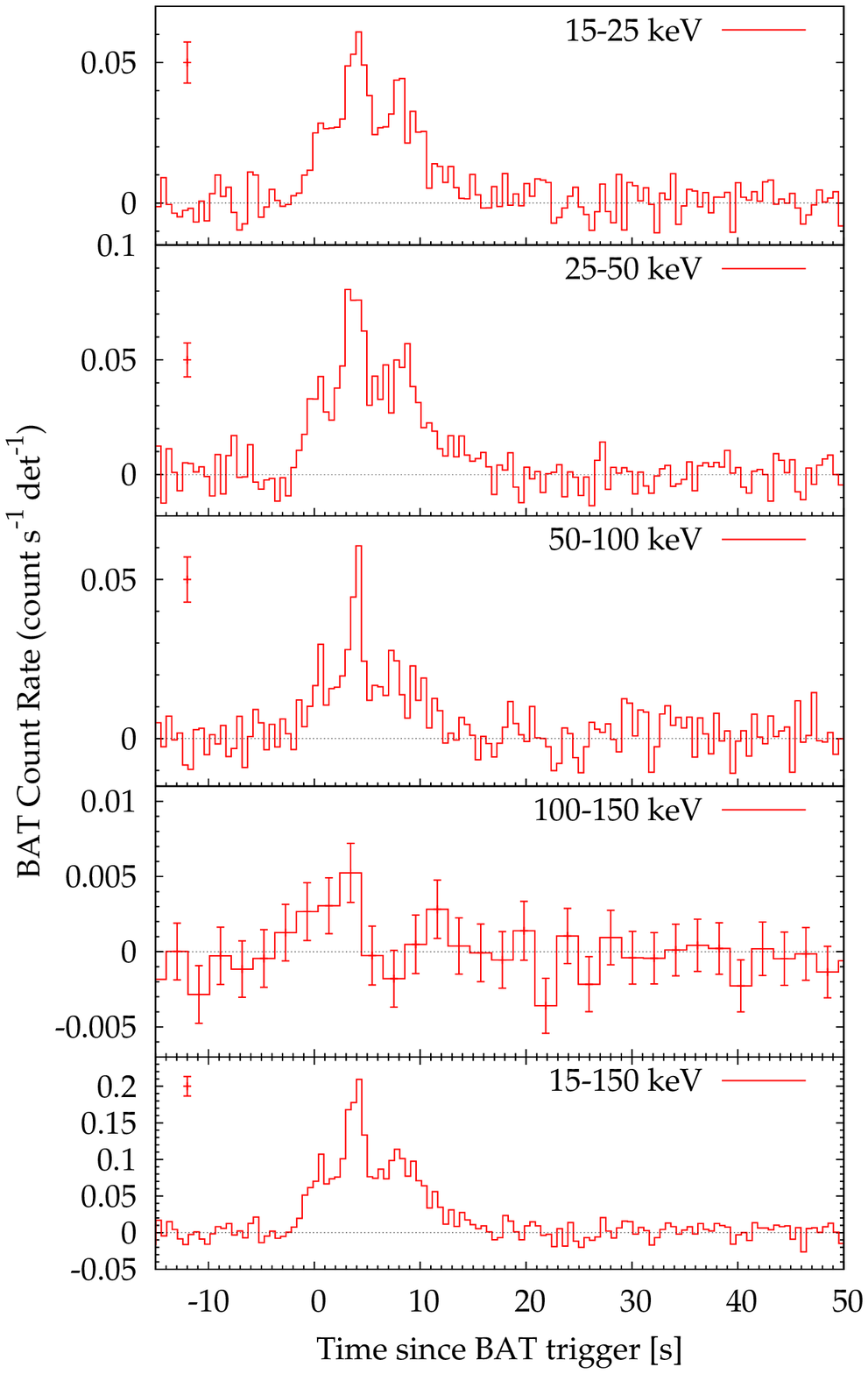,height=14cm}}
\end{center}
\end{figure*}
\begin{singlespace}
\noindent
{\bf Supplementary Fig.~1.} {\footnotesize {\bf BAT mask-weighted light curve.} Four channels and 
combined 0.512 s mask-weighted light curve. 
The light curve of the 100-150 keV energy channel shows a weak
signal, because of the soft spectrum; the corresponding integration time is 2.048 s. Errors are at 1$\sigma$ level.}
\end{singlespace}
\clearpage

\begin{figure*}[t]
\begin{center}
\centerline{\psfig{figure=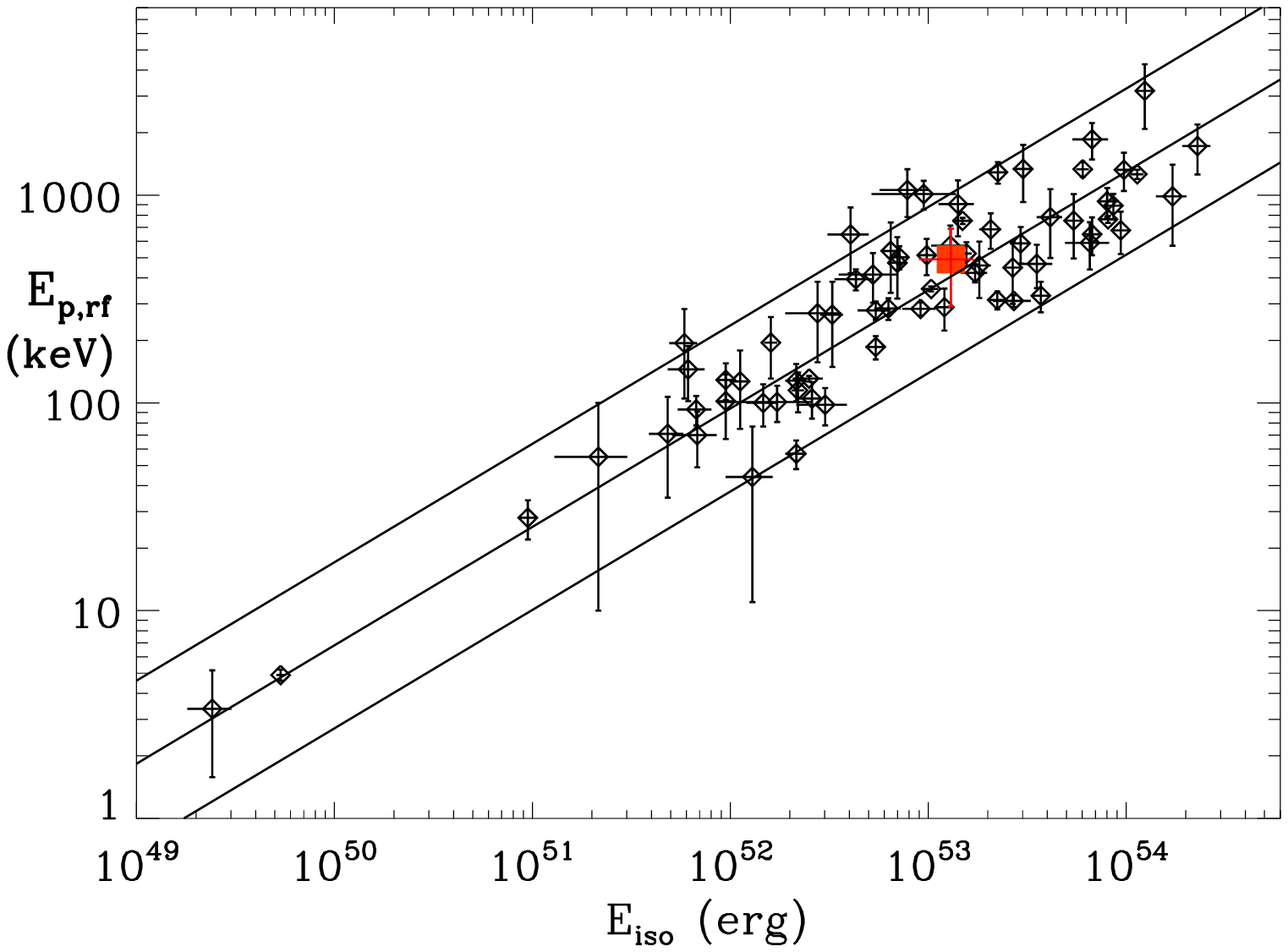,height=11cm}}
\end{center}
\end{figure*}
\begin{singlespace}
\noindent
{\bf  Supplementary Fig.~2. } {\footnotesize {\bf Isotropic energy and peak energy correlation.} Position of GRB~090423 in the $E_{\rm p,rf}$ -- $E_{iso}$  plane based on Swift/BAT$^{4}$ and Fermi/GBM$^{10}$ (fitted with the Band function) results. The lines show the best--fit power--law and the $\pm 2\sigma$ region of the correlation as derived by $^{13}$. Also shown are the 70 GRBs included in the sample analyzed in that work (errors on individual bursts are at 1$\sigma$ level). Given that short GRBs do not follow the
correlation$^{13}$, this evidence supports 
the hypothesis that, despite its cosmological rest--frame duration of 
$\sim 1.3$ s, GRB\,090423 belongs to the long GRB class.}
\end{singlespace}
\clearpage

\begin{figure*}[t]
\begin{center}
\centerline{\psfig{figure=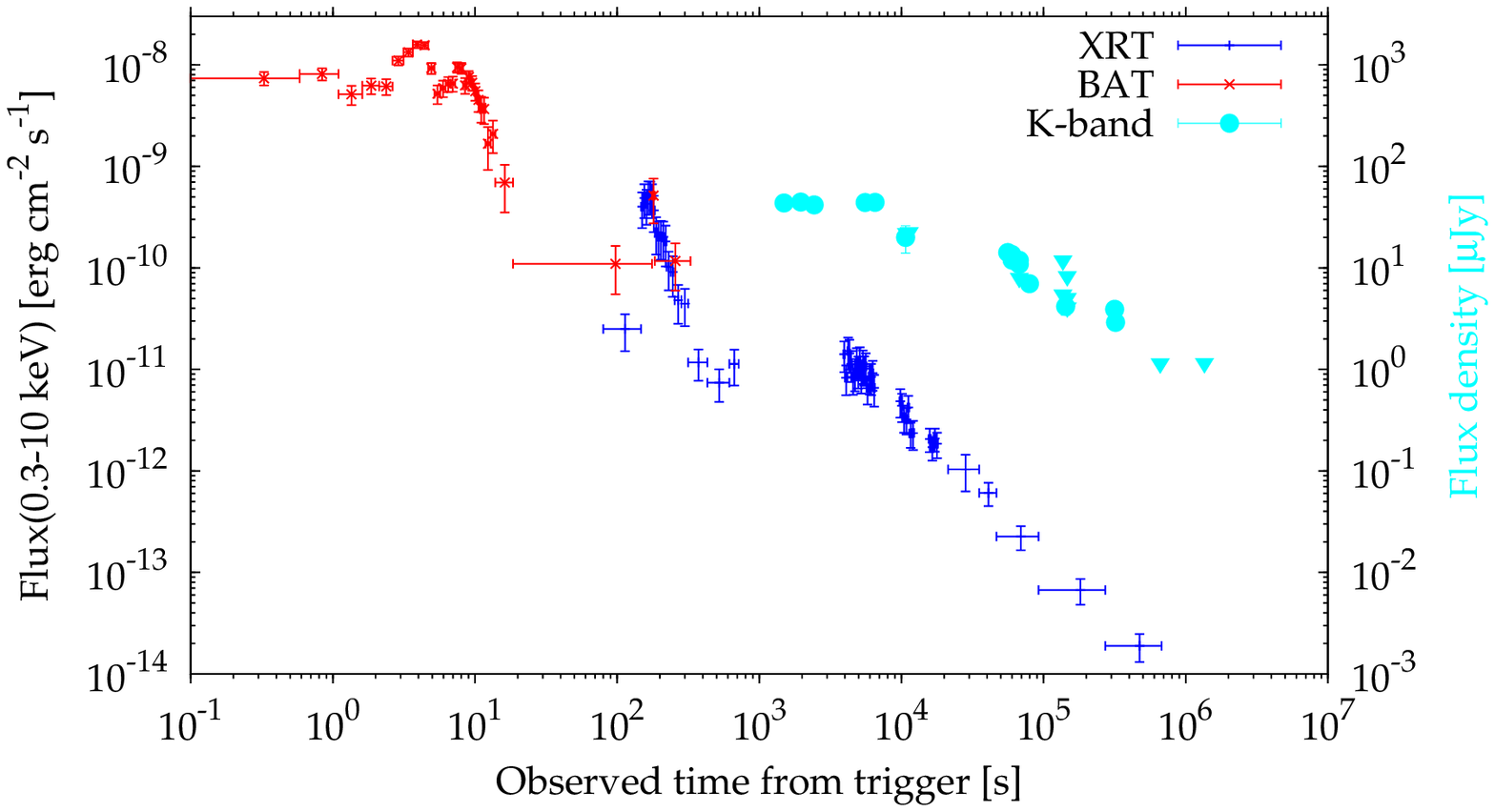,height=11cm}}
\end{center}
\end{figure*}
\begin{singlespace}
\noindent
{\bf Supplementary Fig.~3.} {\footnotesize {\bf Observed light curve.}
Light curve of GRB~090423 as observed by
Swift/BAT (red crosses), Swift/XRT (blue plus) and in the NIR (cyan points).
Errors on fluxes are at 1$\sigma$ level and horizontal bars refer 
to the integration time interval.
The XRT 0.3--10 keV light-curve, starting at 73 s after the burst, 
shows a prominent flare at $t \sim 170$ s (also detected by BAT), and
a flat phase ($\alpha_{X,1}=0.13 \pm 0.11$) followed by a 
rather typical decay (starting at $t=4513\pm491$ s) with power-law index 
$\alpha_{X,2}=1.3\pm0.1$. 
Available photometric data are plotted in the K band (AB magnitude) by 
transforming the fluxes, when 
the observations have been taken in a different filter,  using a 
power law with $\beta = 0.4$, as estimated from the NIR spectral energy
distribution.
A small displacement in time for
contemporary data in different bands is applied in order to increase the
visibility. 
The NIR light curve is consistent with a plateau phase ($t\sim 10^2-10^3$ s) 
followed by a decay with $\alpha_{O}\sim 0.5$ ($t\sim 10^3-10^5$ s).
This decay phase is shallower than the X-ray decay in the same time interval. 
Triangles at $t\sim 10^5$ s report NIR upper limits as obtained by
our second epoch TNG observation with the NICS camera in the Y and J band
and by GROND in the JHK band. These limits are consistent with the temporal 
decay observed by XRT.}
\end{singlespace}
\clearpage

\begin{figure*}[t]
\begin{center}
\centerline{\psfig{figure=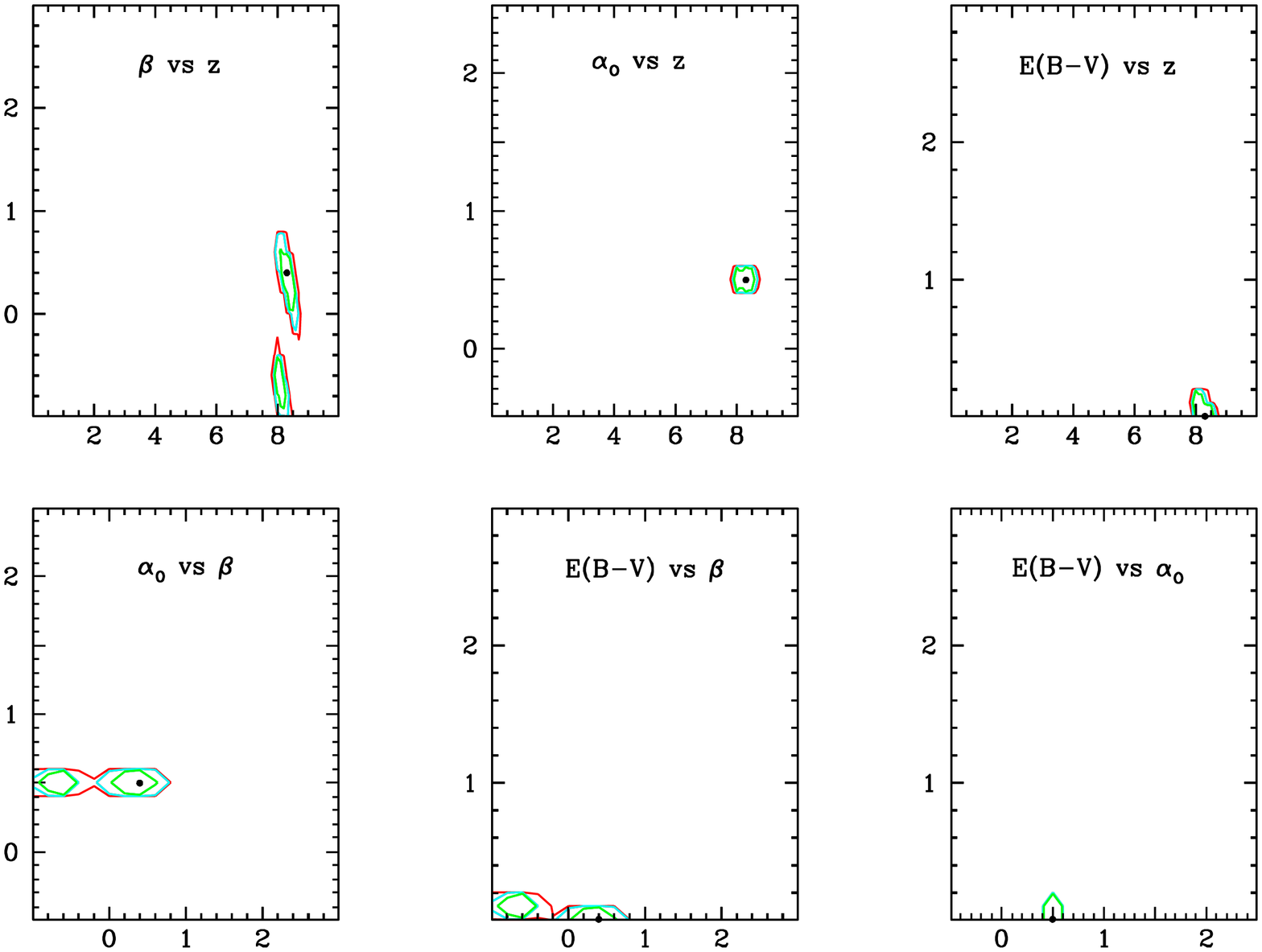,height=15cm,angle=-90}}
\end{center}
\end{figure*}
\begin{singlespace}
\noindent
{\bf  Supplementary Fig.~4. } {\footnotesize {\bf Multi-parameter analysis of the photometric data.} Analysis of available photometric data for GRB090423 in the interval 70 min $< t <$ 1100 min. The code fits a model function with temporal index $\alpha_O$ and spectral index $\beta$, dust extinction $E(B-V)$, and redshift $z$. The different panels show the projection of the four-dimensional confidence intervals on the different two-dimensional planes of interest. The best-fit is marked by the black dot, with the red, cyan, and green contours defining respectively the 68\%, 95\%, and 99.5\% confidence areas. The apparently bimodal distribution in the $\beta$ direction is an artifact of the parameter space discretization.}
\end{singlespace}
\clearpage

\begin{figure*}[t]
\begin{center}
\centerline{\psfig{figure=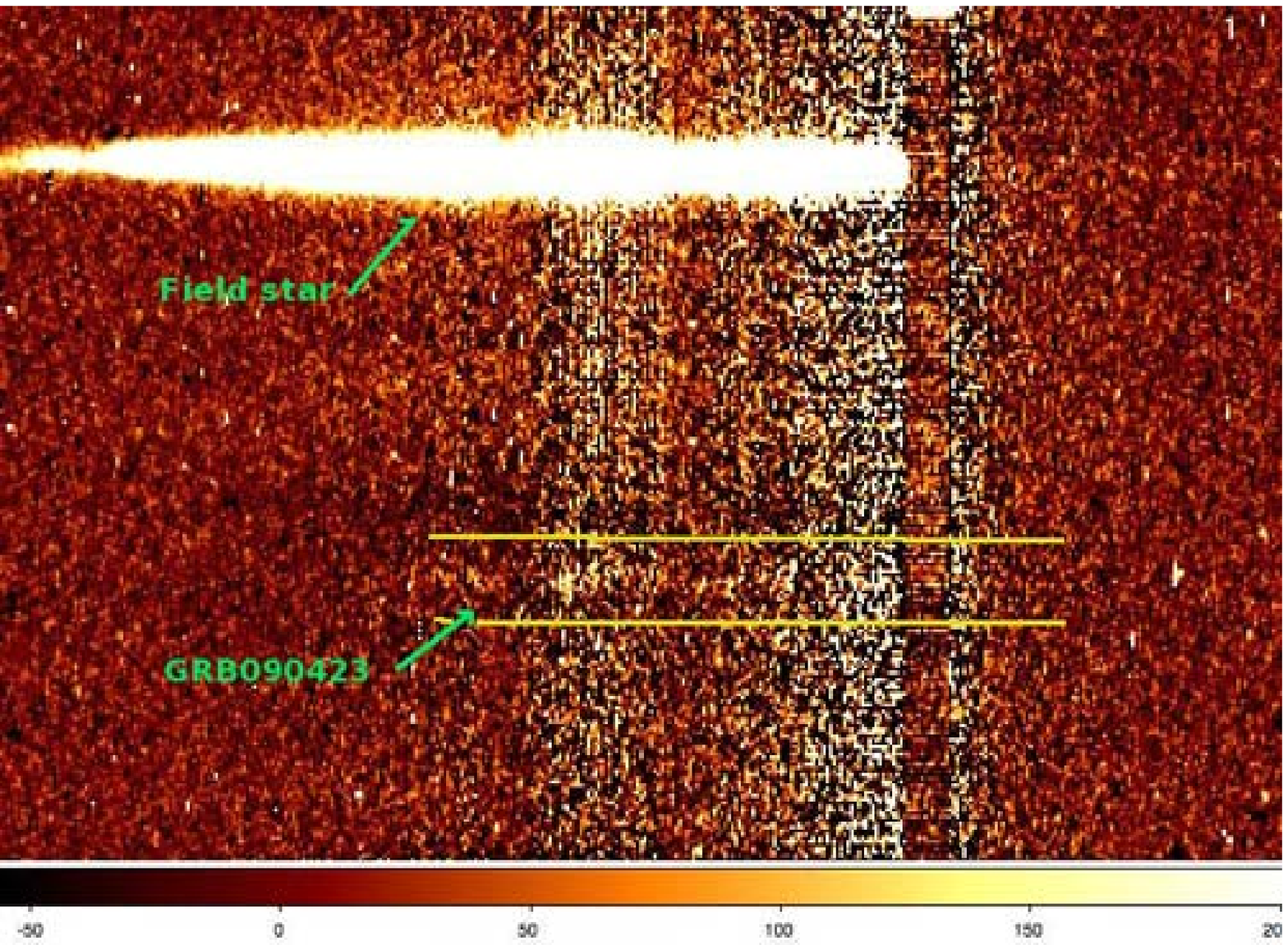,height=9cm}}
\end{center}
\end{figure*}
\begin{singlespace}
\noindent
{\bf  Supplementary Fig.~5. } {\footnotesize {\bf TNG 2-dimensional spectrum.} 
The spectrum has been taken by
$\sim 14$ hrs from the trigger. The spectrum of the nearby reference 
star is also  shown.}
\end{singlespace}
\clearpage

\begin{figure*}[t]
\begin{center}
\centerline{\psfig{figure=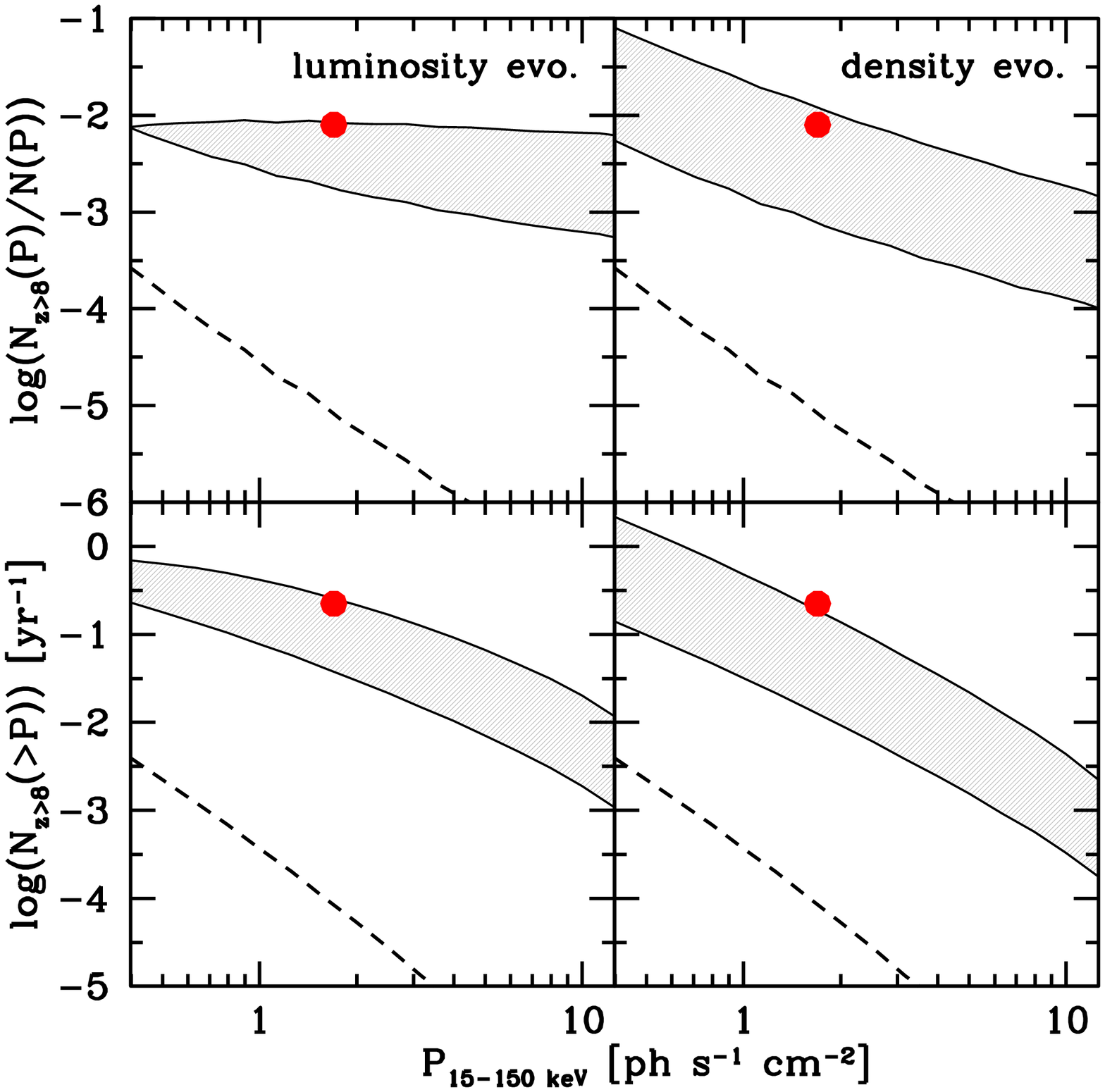,height=11cm}}
\end{center}
\end{figure*}
\begin{singlespace}
\noindent
{\bf  Supplementary Fig.~6.} {\footnotesize {\bf Probability of the occurance of GRB~090423.} Top panels: probability for a GRB with peak photon flux $P$ to be detected by Swift at $z\ge 8$. Luminosity evolution models are shown in the left
panel, where shaded area refers to a typical burst luminosity
increasing as $L_{\rm cut}\propto (1+z)^\delta$ with $\delta=1.5-3$. Density evolution
models are shown in the right panel, where shaded area refers to a
metallicity threshold for GRB formation $Z_{th}=0.02-0.2\;Z_\odot$ (the
lower bound refers to the higher $Z_{th}$).
In both panels, the dashed line shows the no evolution case. The red
point marks the position of GRB~090423. We note that the point represents
a lower limit on the number of detection at $z>8$ since a few very high-$z$ 
bursts may be hidden among those bursts that lack of an optical detection. 
Bottom panels: cumulative
number of GRBs at $z>8$ to be detected by Swift with photon flux
larger than $P$ in one year of Swift observations.}
\end{singlespace}
\clearpage


\begin{thebibliography}{10}


\bibitem[<1>]{Greiner08}
Greiner, J. {\it et al.} GRB~080913 at redshift 6.7. \ApJ {\bf 693},
1610-1620 (2009).

\bibitem[<2>]{Iye06}
Iye, M. {\it et al.} A galaxy at a redshift $z=6.96$.
{\it Nature} {\bf 443}, 186-188 (2006)

\bibitem[<3>]{Komatsu09}
Komatsu, E. {\it et al.}, Five-Year Wilkinson Microwave Anisotropy Probe
Observations: Cosmological Interpretation. \ApJS {\bf 180}, 330-376 (2009).

\bibitem[<4>]{Palmer09}
Palmer, D.M. {\it et al.} GRB 090423: Swift-BAT refined analysis.
{\it GCN Circ.} 9204 (2009).


\bibitem[<5>]{Stratta09}
Stratta, G. \& Perri, M. GRB 090423: Swift-XRT refined analysis.
{\it GCN Circ.} 9212 (2009).

\bibitem[<6>]{Pasquale09}
De Pasquale, M. \& Krimm, H. GRB090423 - Swift/UVOT upper limits.
{\it GCN Circ.} 9210 (2009).


\bibitem[<7>]{Tanvir09}
Tanvir, N. {\it et al.}, A glimpse of the end of the dark ages: the $\gamma$-ray burst of 23 April 2009 at redshift 8.3. {\it Nature} submitted (2009).

\bibitem[<8>]{Thoene09}
 Thoene, C.C. {\it et al.} GRB 090423: TNG Amici spectrum.
{\it GCN Circ.} 9216 (2009).


\bibitem[<9>]{Soto09}
 Fern{\'a}ndez-Soto, A. {\it et al.} GRB 090423: Refined TNG analysis.
{\it GCN Circ.} 9222 (2009).


\bibitem[<10>]{Kienlin09}
von Kienlin, A. GRB 090423: Fermi GBM observation.
{\it GCN Circ.} 9229 (2009).


\bibitem[<11>]{Meszaros}
M{\'e}sz{\'a}ros, P. Gamma-ray bursts. {\it Rep. Prog. Phys.} {\bf 69}, 2259-2322 (2006).

\bibitem[<12>]{Zhang}
Zhang, B. Gamma-Ray Bursts in the Swift Era. {\it Chin. J. Astron. Astrophys.} {\bf 7}, 1-50 (2007).

\bibitem[<13>]{Amati08}
Amati, L. {\it et~al.} On the consistency of peculiar GRBs 060218 and 060614 with the $E_{p,i}-E_{iso}$ correlation. \AaA {\bf 463}, 913-919 (2007). 

\bibitem[<14>]{Schady07}
Schady, P. {\it et al.} Dust and gas in the local environments of gamma-ray bursts. \MNRAS {\bf 377}, 273-284 (2007)

\bibitem[<15>]{Stratta07}
Stratta, G. {\it et al.} Dust Properties at $z=6.3$ in the Host Galaxy of GRB~050904. \ApJL {\bf 661}, 9-12 (2007).

\bibitem[<16>]{Schneider02}
Schneider, R. {\it et al.} First Stars, Very Massive Black Holes, and Metals.
\ApJ {\bf 571}, 30-39 (2002).

\bibitem[<17>]{Springel05}
Springel, V. {\it et al.} Simulations of the formation, evolution and clustering of galaxies and quasars. {\it Nature} {\bf 435}, 629-636 (2005).

\bibitem[<18>]{Nagamine06}
Nagamine, K. {\it et al. } Tracing early structure formation with massive starburst galaxies and their implications for reionization. {\it New Astron.} {\bf 50}, 29-34 (2006).


\bibitem[<19>]{Choudhury08}
Choudhury, T.R., Ferrara, A., Gallerani, S. On the minimum mass of reionization sources. \MNRAS {\bf 385} L58-L62 (2008).

\bibitem[<20>]{Fruchter}
Fruchter, A.S. {\it et al.} Long $\gamma$-ray bursts and core-collapse supernovae have different environments. {\it Nature} {\bf 7092}, 463-468 (2006).


\bibitem[<21>]{Lamb}
Lamb, D.Q., \& Reichart, D.E. Gamma-Ray Bursts as a Probe of the Very High Redshift Universe. \ApJ {\bf 536}, 1-18 (2000). 

\bibitem[<22>]{Guetta05}
Guetta, D., Piran, T., Waxman, E. The Luminosity and Angular Distributions of Long-Duration Gamma-Ray Bursts. \ApJ {\bf 619}, 412-419 (2005)


\bibitem[<23>]{Bromm06}
Bromm, V. \& Loeb, A. High-Redshift Gamma-Ray Bursts from Population III Progenitors.  \ApJ {\bf 642}, 382-388 (2006).

\bibitem[<24>]{SC07}
Salvaterra, R. \& Chincarini, G. The Gamma-Ray Burst Luminosity Function in
the light of the Swift 2 year data. \ApJL {\bf 656}, 49-52 (2007).

\bibitem[<25>]{WB}
Woosley, S. E. \& Bloom, J. S. The Supernova Gamma-Ray Burst Connection. 
{\it Annu. Rev. Astron. Astrophys.} {\bf 44}, 507-556 (2006).

\bibitem[<26>]{RS08}
Salvaterra, R. {\it et al.} Evidence for Luminosity Evolution of Long Gamma-ray Bursts in Swift Data. \MNRAS {\bf 396}, 299-303 (2009).


\bibitem[<27>]{chary}
Chary, R.-R.  The Stellar Initial Mass Function at the Epoch of Reionization.
\ApJ {\bf 680}, 32-40 (2008).


\bibitem[<28>]{Bolton07}
Bolton, J.S. \& Haehnelt, M.G. The observed ionization rate of the intergalactic medium and the ionizing emissivity at $z\ge 5$: evidence for a photon-starved and extended epoch of reionization. \MNRAS {\bf 382}, 325-341 (2007).


\bibitem[<29>]{Furlanetto09}
Furlanetto, S.R. \& Mesinger, A. The ionizing background at the end of reionization. \MNRAS {\bf 394}, 1667-1673 (2009).

\bibitem[<30>]{Stiavelli}
Stiavelli, M. {\it From First Light to Reionization: The End of the Dark Ages.} 
Vch Verlagsgesellschaft Mbh. (2009).




\end{thebibliography}

\bigskip
\bigskip

\begin{thebibliography}{10}


\bibitem[<31>]{Band93}
Band, D.L. {\it et al.} BATSE observations of gamma-ray burst spectra. I - Spectral diversity. \ApJ {\bf 413}, 281-292 (1993).

\bibitem[<32>]{Zhang09}
Zhang, B.-B. \& Zhang, B. GRB 090423:  pseudo burst at z=1 and its relation to GRB 080913. {\it GCN Circ.} 9216 (2009).

\bibitem[<33>]{Amati08}
Amati, L. {\it et~al.} On the consistency of peculiar GRBs 060218 and 060614 with the $E_{p,i}-E_{iso}$ correlation. \AaA {\bf 463}, 913-919 (2007). 

\bibitem[<34>]{Norris05}
Norris J.P. {\it et al.} Long-Lag, Wide-Pulse Gamma-Ray Bursts.
\ApJ {\bf 627}, 324-345 (2005).

\bibitem[<35>]{Dickey90}
Dickey, J.M. \& Lockman F.J. H I in the Galaxy. {\it Annu. Rev. Astron. Astrophys.} {\bf 28}, 215-261 (1990). 

\bibitem[<36>]{Kalberla04}
Kalberla, P.M.W. {\it et al.} A New Whole HI Sky Survey. Proceedings of ASP Conference {\bf 317}, 13 (2004).

\bibitem[<37>]{Evans09}
ev. sostituire con Campana in prep. Evans, P.A. {\it et al.} Methods and results of an automatic analysis of a complete sample of Swift-XRT observations of GRBs. ArXiv e-print:0812.3662 (2008).

\bibitem[<38>]{Campana07}
Campana, S. {\it et al.} A Metal-rich Molecular Cloud Surrounds GRB 050904 at Redshift 6.3. \ApJL {\bf 654}, 17-20 (2007).

\bibitem[<39>]{Soto03}
Fern{\'a}ndez-Soto A. {\it et al.} z-ph-REM: A photometric redshift code for the REM telescope. arXiv:astro-ph/0309492 (2003).

\bibitem[<40>]{Baffa01}
Baffa, C. {\it et al.} NICS: The TNG Near Infrared Camera Spectrometer. \AaA {\bf 378}, 722-728 (2001).

\bibitem[<41>]{Oliva00}
Oliva E. Infrared instrumentation for large telescopes : an alternative approach. MemSAIt {\bf 71}, 861 (2000)

\bibitem[<42>]{Hopkins}
Hopkins, A. M. \& Beacom, J. F. On the Normalization of the Cosmic Star Formation History. \ApJ {\bf 651}, 142-154 (2006).

\bibitem[<43>]{Langer06}
Langer, L. \& Norman, C. A. On the Collapsar Model of Long Gamma-Ray Bursts:Constraints from Cosmic Metallicity Evolution. \ApJL {\bf 638}, 63-66 (2006).

\bibitem[<44>]{Natarajan05}
Natarajan, P. {\it et al.} The redshift distribution of gamma-ray bursts revisited. \MNRAS {\bf 364}, L8-L12 (2005).

\bibitem[<45>]{Preece00}
Preece R. D. {\it et al.} The BATSE Gamma-Ray Burst Spectral Catalog. I. High Time Resolution Spectroscopy of Bright Bursts Using High Energy Resolution Data. {\it Astrophys. J. Suppl. Ser.} {\bf 136}, 19-36 (2000).

\bibitem[<46>]{Yonetoku04}
Yonetoku D. {\it et al.} Gamma-Ray Burst Formation Rate Inferred from the Spectral Peak Energy-Peak Luminosity Relation. \ApJ {\bf 609}, 935-951 (2004).

\bibitem[<47>]{PM01}
Porciani, C. \& Madau, P. On the Association of Gamma-Ray Bursts with Massive Stars: Implications for Number Counts and Lensing Statistics. \ApJ {\bf 548}, 522-531 (2001).

\end{thebibliography}
\end{document}